\documentclass{sig-alternate}

\usepackage{acronym}
\usepackage{subfigure}
\usepackage{booktabs}
\usepackage{algorithm}
\usepackage{algcompatible}
\usepackage{url}
\usepackage{color}

\acrodef{MIP}{Mixed Integer Program\-ming}
\acrodef{P2P}{Peer-to-Peer}

\def\sharedaffiliation{%
\end{tabular}
\begin{tabular}{c}}
\makeatletter
\let\@copyrightspace\relax
\makeatother
\begin{document}

\title{Auction-Based Resource Allocation in Digital Ecosystems\footnotemark}

\numberofauthors{3}
\author{
\alignauthor Moreno Marzolla\\
\email{marzolla@cs.unibo.it}
\alignauthor Stefano Ferretti\\
\email{sferrett@cs.unibo.it}
\alignauthor Gabriele D'Angelo\\
\email{g.dangelo@unibo.it}
\sharedaffiliation
\affaddr{Dipartimento di Scienze dell'Informazione}\\
\affaddr{Universit\`a di Bologna}\\
\affaddr{Mura A. Zamboni 7, I-40127 Bologna, Italy}\\
}

\toappear{}
\maketitle

\footnotetext{The publisher version of this paper is available at \url{http://dx.doi.org/10.1109/Mobilware.2013.16}.
\textbf{{\color{red}Please cite this paper as: Moreno Marzolla, Stefano Ferretti, Gabriele D'Angelo. Auction-Based Resource Allocation in Digital Ecosystems. Proceedings of the 6th International Conference on MOBILe Wireless MiddleWARE, Operating Systems, and Applications (MobilWare 2013), Bologna (Italy), November, 2013.}}}

\begin{abstract}
The proliferation of portable devices (PDAs, smartphones, digital
multimedia players, and so forth) allows mobile users to carry around
a pool of computing, storage and communication resources. Sharing
these resources with other users (``Digital Organisms'' -- DOs) opens
the door to novel interesting scenarios, where people trade resources
to allow the execution, anytime and anywhere, of applications that
require a mix of capabilities. In this paper we present a fully
distributed approach for resource sharing among multiple devices owned
by different mobile users. Our scheme enables DOs to trade
computing/networking facilities through an auction-based mechanism,
without the need of a central control. We use a set of numerical
experiments to compare our approach with an optimal (centralized)
allocation strategy that, given the set of resource demands and
offers, maximizes the number of matches. Results confirm the
effectiveness of our approach since it produces a fair allocation of
resources with low computational cost, providing DOs with the means to
form an altruistic digital ecosystem.
\end{abstract}


%%%%%%%%%%%%%%%%%%%%%%%%%%%%%%%%%%%%%%%%%%%%%%%%%%%%%%%%%%%%%%%%%%%%%%%%%%%%%%
\section{Introduction}

Mobile users are evolving: while in the recent past people used their
mobile devices just for ``simple'' tasks such as checking e-mail or
browsing the Web, the rise of novel social applications fosters a
massive use of ubiquitous services. Current Operating Systems for
mobile devices (e.g.,~Android, iOS) allow the execution of
applications that, for instance, publish user's geographical position
and other context-aware information on social networking
services. However, these forms of interaction are usually based on the
classic client-server approach, i.e.,~the mobile device connects to a
central service through its own Internet connection. The proliferation
of heterogeneous devices with different capabilities (computation,
communication, data storage, sensors and actuators) gives rise to new
scenarios that promote the cooperation among individuals in order to
guarantee the provision of ``always on'' services~\cite{iwcmc}.

As an example, consider a medical doctor who receives an urgent call
while attending a meeting with other colleagues. He needs to check
some medical data to diagnose a particular illness; unfortunately, the
tablet PC he carried with him has not enough memory and computational
power to execute the job. Therefore, he ``rents'' CPU power from one
of his colleagues high-end laptop to carry on the analysis.

As another example, consider a user that wants to upload some pictures
made with her smart-phone/camera on Flickr or on the wall of her
social Web application, but she is not provided with network
connectivity. Hence she exploits the 3G mobile connection of a
neighboring friend using that as a gateway to the net via an ad-hoc
short range connection (e.g.,~Bluetooth).

In general, mobile users have many different devices in their pockets
and suitcase, each of them with specific hardware and software
characteristics. Quite often such devices are not enabled for seamless
interaction with other devices belonging to the same owner; sharing
resources among different people is even more challenging. The
possibility for a user to exploit, in a~\ac{P2P} and altruistic way,
computing facilities owned by (known and trusted) neighbors requires
mechanisms for automatic service discovery and negotiation, and for
access control.

A system architecture supporting the scenarios above has been recently
proposed~\cite{iwcmc}. Each mobile user is considered as a ``digital
organism'' (DO), composed by many different devices belonging to the
same human being. Each DO may share resources with peer DOs using
auto-configuration strategies. Then, a community of interacting DOs
can be thought of as a ``digital ecosystem''.  Each DO in the
ecosystem contributes by providing its own unused resources to its
(friend/trusted) neighbors. Therefore, the community of DOs exploits
self-organization protocols and~\ac{P2P} strategies to create a
cooperating, altruistic ecosystem.

Of course, privacy and security issues should be carefully considered,
especially if data are to be distributed to other users. Addressing
these issues in a mobile environment is highly nontrivial, especially
when the remote nodes are completely untrusted. Therefore, we assume
that each DO will preferably connect to other DOs towards which there
is explicit trust (e.g., because users know each other). For example,
a user traveling by train shall be willing to share a network
connection with some (known and trusted) traveling companions.

In this paper we address the problem of optimizing the allocation of
resources in a digital ecosystem, by matching resource requests and
offers. We consider a~\ac{P2P} overlay which connects DOs; each DO may
offer and/or require resources, which are traded with neighboring
DOs. We present a fully distributed scheme that can match demands and
offers, allowing resources to be provisioned efficiently and
fairly. We use a market-based approach in which requests are handled
through ascending clock auctions~\cite{Ausubel2004}. We assume that
each DO can use some form of ``virtual currency'' (tokens) as a form
of payment for resources usage: this allows a fair allocation that
balances supply and demand~\cite{Kelly04,stokely}. First, we describe
a distributed algorithm to carry on the auction; then, we formulate
the resource allocation problem as a~\ac{MIP} optimization problem,
which is used to compute the maximum number of requests that can be
matched. We compare the optimal allocation with the one provided by
our our cheap, local strategy. Results show that our approach produces
good allocations and requires low computational cost: this makes the
auction-based allocation strategy particularly appropriate for sharing
resources among devices with very limited computational power.

The rest of this paper is organized as follows. In
Section~\ref{sec:formulation} we give a precise formulation of the
problem we are addressing. Section~\ref{sec:auction} presents our
auction-based resource allocation scheme. In Section~\ref{sec:exp} we
discuss numerical results obtained from a set of synthetic simulation
experiments. In Section~\ref{sec:related} we revise the literature and
contrast our approach with some related work. Finally, concluding
remarks are provided in Section~\ref{sec:conc}. Additional details on
the auction algorithm, and the~\ac{MIP} formulation of the
optimization problem, are given in the Appendix.

%%%%%%%%%%%%%%%%%%%%%%%%%%%%%%%%%%%%%%%%%%%%%%%%%%%%%%%%%%%%%%%%%%%%%%%%%%%%%%
\section{Problem Formulation}\label{sec:formulation}

We consider a set $\mathbf{R} = \{1, \ldots, R\}$ of $R$ different
resource types (e.g., network connectivity, processing power, storage,
and so on).  $\mathbf{N} = \{1, \ldots, N\}$ denotes a set of $N$
users trading these resources: each user can be either a \emph{buyer}
(if he requests resources) or a \emph{seller} (if he offers
resources). The same user may play the role of buyer and seller at the
same time, offering surplus resources while buying those he needs.

For each user $i \in \mathbf{N}$, we denote with $\mathit{Req}_{ri}$
the amount of type $r$ resource requested by $i$; for each $j \in
\mathbf{N}$, we denote with $\mathit{Off}_{rj}$ the amount of type $r$
resource offered by $j$; quantities are not restricted to be integer.
The vectors $\mathit{Req}_{\bullet i}$ and $\mathit{Off}_{\bullet j}$
are called \emph{resource bundles}\footnote{We use the symbol
  $\bullet$ as a shortcut to denote a slice of a multi-dimensional
  vector; therefore, $\mathit{Req}_{\bullet i}$ denotes the slice
  $(\mathit{Req}_{1 i},\mathit{Req}_{2 i}, \ldots, \mathit{Req}_{R
    i})$}.

As an example, if there are two resource types (``CPU'' and ``Network
Bandwidth''), then a resource bundle $(0.1\ \text{MIPS},
\\ 200\ \mathrm{KB}/s)$ can be interpreted as a request (or an offer)
for $0.1$ MIPS of CPU power and $200$ $\mathrm{KB}/s$ of network
bandwidth. Unit of measures will be omitted in the following.

We assume that each user (node) is equipped with some form of wireless
connectivity which enables short range interaction with a set of
neighbors using an ad-hoc network topology. We model this with an $N
\times N$ adjacency matrix $M_{ij}$, where users $i$ and $j$ can
interact iff $M_{ij} = M_{ji} = 1$; matrix $M_{ij}$ is symmetric, so
that interactions are always bidirectional.

Each user can get resources from, or provide resources to, one of his
direct neighbors; multi-hop interactions are not allowed. Multi-hop
interactions would be much harder to handle, since appropriate routing
strategies should be employed to ensure connectivity in spite of
individual users moving and losing contact with neighbors.  We
introduce the binary decision variable $X_{irj}$ which equals $1$ iff
user $i$ obtains resource $r$ from user $j$. If $X_{irj} = 1$, then
$i$ must obtain exactly $\mathit{Req}_{ri}$ items of resource $r$ from
$j$. The allocation $X_{irj}$ must satisfy the following constraints:

\begin{enumerate}
\item Each buyer $i$ must obtain the requested quantities
  $\mathit{Req}_{ri}$ of all resources $r$ in his bundle, or none at
  all. Partially fulfilled requests are not allowed.
\item For each $r$, the requested quantity $\mathit{Req}_{ri}$
  must be provided by a single seller $j$ (if the request is
  satisfied at all).
\item For each $r$, the offered quantity $\mathit{Off}_{rj}$ can be
  fractioned across multiple buyers (i.e., a seller is not forced to
  provide all items of resource $r$ to a single buyer).
\item If user $i$ gets resource $r$ from $j$, then the amount
  requested by $i$ must not exceed the amount offered by $j$:
  $\mathit{Req}_{ri} \leq \mathit{Off}_{rj}$.
\item For all $r \in \mathbf{R}$, $\sum_{i \in \mathbf{N}}
  \mathit{Req}_{ri} X_{irj} \leq \mathit{Off}_{rj}$, where the
  left-hand side represents the total amount of resource $r$ provided
  by seller $j$. This means that the total amount of resource $r$
  provided by $j$ to all buyers must not exceed his capacity.
\item $X_{irj}$ can be $1$ only if $M_{ij} = 1$: interactions are
  only allowed between neighbors.
\end{enumerate}

\begin{table}[t]
\centering%
\begin{tabular}{rp{.75\columnwidth}}
\toprule
$\mathbf{N} :=$ & $\{1, \ldots, N\}$ set of users\\
$\mathbf{R} :=$ & $\{1, \ldots, R\}$ set of resource types\\
$\mathit{Req}_{ri} :=$ & Amount of resource $r$ requested by user $i$\\
$\mathit{Off}_{rj} :=$ & Amount of resource $r$ offered by user $j$\\
$\mathit{RP}_i :=$ & Reserve price of buyer $i$: maximum amount $i$ is willing to pay for his requested bundle\\
$\mathit{SP}_{rj} :=$ & Unitary selling price of resource $r$ offered by user $j$\\
$M_{ij} :=$ & $1$ iff $i$ can interact with $j$\\
$X_{irj} :=$ & 1 iff $i$ obtains resource type $r$ from $j$\\
\bottomrule
\end{tabular}
\caption{Notation used in this paper}\label{tab:notation}
\end{table}

The notation used in this paper is summarized in
Table~\ref{tab:notation} (additional symbols shown in the table will
be introduced in the next section).

The problem of finding an optimal allocation $X_{irj}$ which maximizes
the number of matched requests (i.e., maximizing the number of
requests which can be satisfied by some seller) can be formulated as
a~\ac{MIP} optimization problem (details are given in
Appendix~\ref{app:mip}). However, solving the optimization problem is
impractical for several reasons: (i) global knowledge of all
parameters is required, whereas each peer has only local knowledge;
(ii) solving large instances of the optimization problem using
general-purpose~\ac{MIP} solvers is time cosuming; (iii) the optimal
allocation may not even be desirable, since the constraints above do
not take into account any measure of fairness between users. The lack
of fairness is a parti\-cu\-larly serious limitation, since it gives
users no incentive to share their resources. While it would be
possible to extend the optimization problem to take fairness into
account, the other limitations would still apply.

In the next section we propose a distributed algorithm for binding
resource requests with resource availability; our algorithm is fully
decentralized and uses an economic model based on ascending auctions;
``virtual currency'' is used to compensate transactions and stop free
riders.

\section{Auction-Based Resource Allocation}\label{sec:auction}

We propose a fully distributed algorithm to compute a fair allocation
of the resources offered by sellers.  The algorithm is lightweight and
fully decentralized, since it will be executed on portable devices
(smartphones, PDAs, notebooks) connected through an ad-hoc network
infrastructure.

Our algorithm is based on ascending clock auctions~\cite{Ausubel2004};
in this type of auctions, the auctioneer defines the price of goods,
and the bidders reply with the quantities they want to buy. In case of
excess demand, the auctioneer raises the price and calls for new bids.
This mechanism is iterated until there is no excess demand. Successful
bidders pay the last announced price. In our scenario, each seller
engages an auction with all potential buyers and adjusts the resource
prices independently of other sellers.

\begin{figure}[t]
\centering\includegraphics[scale=.8]{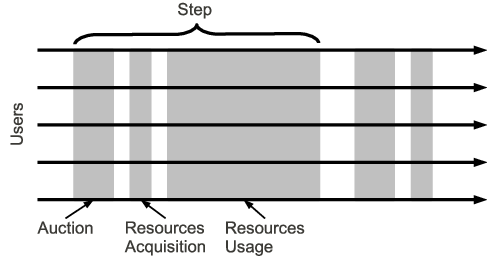}
\caption{Auction-based resource allocation}\label{fig:auction}
\end{figure}

Resource allocation works as a sequence of steps, as depicted in
Figure~\ref{fig:auction}. Each step is divided into three phases.
During the auction phase, all users engage an auction which consists
of multiple rounds; during each round, sellers broadcast the unitary
prices and quantities of the resources they offer, and potential
buyers place bids. The sellers which receive excess demand raise the
selling price and start a new round. At the end of the auction,
successful buyers pay their dues to the sellers and can access the
provided resources. At this point, usage is granted for some amount of
time, after which a new step is initiated.

Each user has some amount of virtual currency (tokens), which can be
spent to acquire resources provided by other peers, or earned by
selling access to his resources. In the following we assume that an
adequate scheme exists to employ virtual concurrency; dealing with the
details is out of the scope of this paper, since several proposals
exist such as those described
in~\cite{nair,Sirivianos:2007,Wei:2006,Yang:2003}.

Each node $i$ autonomously computes a slice of the allocation matrix
$X_{irj}$, i.e., decides from which sellers to buy the resources in
his request bundle. Each seller $j$ broadcasts the vector
$\mathit{Off}_{\bullet i} = \left( \mathit{Off}_{1j}, \ldots,
\mathit{Off}_{Rj} \right)$ of the amounts of resources offered,
together with the unitary selling prices $\mathit{SP}_{\bullet j} =
\left( \mathit{SP}_{1j}, \ldots, \mathit{SP}_{Rj} \right)$. Each
seller can define initial prices at the beginning of an auction phase;
such prices are called \emph{reserve prices}, and represent the
minimum prices at which a seller is willing to offer his resources.

Each buyer $i$ can place \emph{bids} to all sellers $j$ in his
neighborhood from which he wants to acquire resources. A bid is a
$R$-dimensional vector with elements $\mathit{Req}_{ri} X_{irj}$. Each
element of the bid is a proposal to acquire $\mathit{Req}_{ri}
X_{irj}$ items of resource $r$ from seller $j$ at the unitary price
$\mathit{SP}_{rj}$. Buyer $i$ places a bid for resource $r$ of $j$
provided that (i) $\mathit{Off}_{rj} \geq \mathit{Req}_{ri}$ (the
amount requested does not exceed the amount offered), and (ii) the
unitary price $\mathit{SP}_{rj}$ of seller $j$ is the minimum over all
potential sellers. Each buyer $i$ tries to acquire resource $r$ from
the seller providing a sufficient quantity, at the lower price.

Since there is a finite amount of each resource type, it is necessary
to handle the situation in which the demand is larger than the supply
provided by a seller. We employ a mechanism based on \emph{ascending
  auctions}, which eventually produces an allocation $X_{irj}$
satisfying the constraints described in Section~\ref{sec:formulation}.

If all bids placed by $i$ were to be accepted from the sellers, the
total cost for the buyer would be
\[
\sum_{r \in \mathbf{R}} \sum_{j \in  \mathbf{N}} \mathit{SP}_{rj} \mathit{Req}_{ri} X_{irj}\label{eq:cost}
\]
Each buyer $i$ has a maximum reserve price $\mathit{RP}_i$, which
represents the total maximum amount of money he is willing to spend
for the bundle $\mathit{Req}_{\bullet i}$\footnote{Note that the
  meaning of \emph{reserve price} is different for buyers and
  sellers.}. Buyer $i$ places bids as long as the cost~\eqref{eq:cost}
is not greater than his reserve price $\mathit{RP}_i$.

Each seller $j$ collects all bids he receives, and replies back to the
buyers with a new vector of (possibly updated) prices $\left(
\mathit{SP}'_{1j}, \ldots, \mathit{SP}'_{Rj}\right)$, where:
\[
\mathit{SP}'_{rj} = \begin{cases}
\mathit{SP}_{rj} & \mbox{if $\sum_{i \in \mathbf{N}} \mathit{Req}_{rj} X_{irj} \leq \mathit{Off}_{rj}$}\\
\mathit{SP}_{rj} + \Delta P & \mbox{otherwise}
\end{cases}
\]
This means that $j$ increases the price by some quantity $\Delta P$ of
each resource $r$ for which there is excess demand. Potential buyers
who placed bids resulting in excess demands must either issue a new
bid at the new prices, or give up. The pseudo-code of the seller and
buyer algorithms are given in Appendix~\ref{app:auction}.

\begin{figure}[t]
\centering\includegraphics[scale=.8]{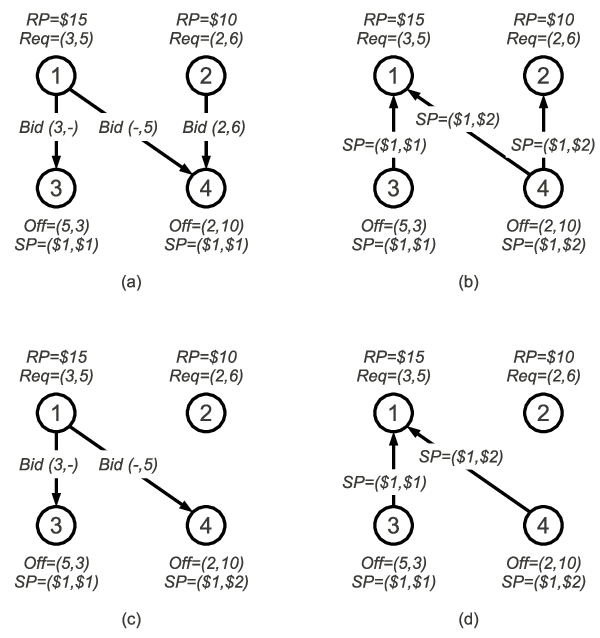}
\caption{Example of resource allocation through ascending
  auction}\label{fig:example}
\end{figure}

Figure~\ref{fig:example} shows a simple example with $N=4$ nodes and
$R=2$ resource types. Nodes $1$ and $2$ are buyers, and can interact
with sellers $\{3, 4\}$, and $4$ respectively. The amounts of
resources requested or offered is shown next to each node: for
example, node $1$ requests $3$ items of resource $1$ and $5$ items of
resource $2$.

After the sellers broadcast the offered quantities and selling prices
(in this case, unitary prices are initially set to $1$), buyers make
their initial bids (Figure~\ref{fig:example}(a)). Specifically, user
$1$ requests $3$ items of resource $1$ to user $3$, and $5$ items of
resource $2$ to user $4$. User $2$ requests $2$ items of resource $1$
and $10$ items of resource $2$ to user $4$. If all bids were accepted,
both nodes $1$ and $2$ would spend $\$8$.

According to the bids above, user $4$ has excess demand on resource
$2$ since $11$ items are requested but only $10$ are
available. Therefore, node $4$ increases the unitary price of resource
$2$ to $\$2$ (we use $\Delta P = \$1$); seller $3$ has no excess
demand, so he replies with the same unitary prices
(Figure~\ref{fig:example}(b)). 

With the new prices, the bundle requested by user $1$ would cost $\$3
\times 1 + \$2 \times 5 = \$13$ (which is below the reserve price
$\mathit{RP}_1 = \$15$), and the requested bundle of user $2$ would
cost $\$14$, which is above the reserve price $\mathit{RP}_2 =
\$10$. Therefore, user $2$ gives up, while user $1$ resubmit his bids
(Figure~\ref{fig:example}(c)). Since there is no excess demands, the
auction ends and user $1$ can finally acquire his bundle
(Figure~\ref{fig:example}(d)).

%%%%%%%%%%%%%%%%%%%%%%%%%%%%%%%%%%%%%%%%%%%%%%%%%%%%%%%%%%%%%%%%%%%%%%%%%%%%%%
\section{Experimental Evaluation}\label{sec:exp}

In this section we analyze the performance of the auction-based
resource allocation algorithm using a set of numerical experiments.
We consider different network sizes (with $N=10, 20, 50$ users,
respectively) and different numbers of resource types ($R=3, 5, 7$).
For each combination of $N$ and $R$ we perform $T=10$ allocation
steps. Each step involves the definition of requested and offered
bundles (see below), and running an auction to match them. At the very
beginning, each user is given a budget of 100 tokens; furthermore,
before starting each step several initializations are performed, as
follows.

First, we generate a random network with link density $0.3$ (this
means that on average, $30\%$ of the elements of the adjacency matrix
$M_{ij}$ are nonzero). $20\%$ of the users are randomly assigned the
role of pure buyers (the offered bundles are set to zero), while the
others are pure sellers (the requested bundles are set to zero). Each
user is also assigned a random demand or offer vector: the number of
items of each resource type that are offered/requested are drawn with
uniform probability from the discrete set $\{1, 2, \ldots, 10\}$. The
initial reserve prices for sellers are set to $1$, and the price
increment is $\Delta P=1$. The reserve price for buyer $i$ is set to
$\epsilon_i \times \sum_{r \in \mathbf{R}} \mathit{Req}_{ri}$, where
$\epsilon_i$ is uniformly chosen in $[1.5, 2]$; $\sum_{r \in
  \mathbf{R}} \mathit{Req}_{ri}$ is the cost of bundle
$\mathit{Req}_{\bullet i}$ when all items have unitary cost. With the
setup above we ensure that each node has sufficient liquidity to
satisfy requests for resources, since each user will act most of the
time as seller. In order to cope with statistical fluctuations, we
executed 20 independent replications of each sequence of $T$ steps; at
the end of all replications, average values and confidence intervals
at $(1- \alpha) = 0.9$ confidence level are computed.

\paragraph*{Number of matches} We first analyze the total number of matches,
i.e., the total number of requests which can be satisfied at the end
of the sequence of $T$ allocation steps. We compare the value obtained
using the auction with the optimal value obtained by matching requests
using the~\ac{MIP} problem on Appendix~\ref{app:mip}; the optimization
problem has been solved using GLPK~\cite{glpk}.

\begin{table}[t]
\centering%
\begin{tabular}{rrrrc}
\toprule
    &     & \multicolumn{2}{c}{Matches} \\
\cmidrule{3-4}
$N$ & $R$ & Auction & Optimal & Auction/Optimal\\
\midrule
10 & 3 & $14.60 \pm 1.48$ & $14.70 \pm 1.45$ & 0.99 \\
20 & 3 & $46.10 \pm 2.33$ & $47.80 \pm 2.41$ & 0.96 \\
50 & 3 & $162.50 \pm 4.55$ & $172.50 \pm 3.14$ & 0.94 \\
\addlinespace
10 & 5 & $10.40 \pm 1.07$ & $10.40 \pm 1.07$ & 1.00 \\
20 & 5 & $33.50 \pm 2.93$ & $34.10 \pm 3.12$ & 0.98 \\
50 & 5 & $149.80 \pm 4.44$ & $161.20 \pm 2.58$ & 0.93 \\
\addlinespace
10 & 7 & $6.30 \pm 1.64$ & $6.30 \pm 1.64$ & 1.00 \\
20 & 7 & $26.40 \pm 3.30$ & $27.80 \pm 3.50$ & 0.95 \\
50 & 7 & $126.80 \pm 6.41$ & $141.70 \pm 5.38$ & 0.89 \\
\bottomrule
\end{tabular}
\caption{Number of matches (higher is better).}\label{tab:num_matches}
\end{table}

\begin{figure}
\centering\includegraphics[width=\columnwidth]{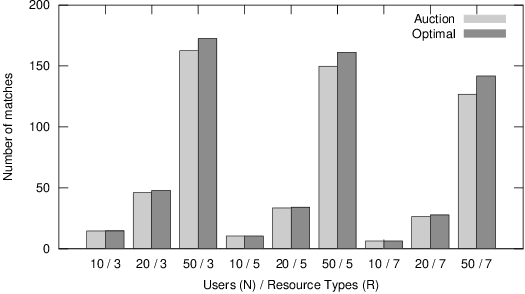}
\caption{Performance results: number of matches (higher is better)}\label{fig:num_matches}
\end{figure}

Raw results are reported in Table~\ref{tab:num_matches}. The column
labeled ``Auction'' shows the total number of matches at the end of
the $T$ auctions, computed with the ascending auction algorithm
proposed in this paper. Column labeled ``Optimal'' shows the maximum
number of matches computed using the optimization problem. The results
has been plotted in Figure~\ref{fig:num_matches}): we can see that the
number of matches produced by the auction algorithm is only slightly
less than the optimum value. It is important to report that for the
larger systems ($N=50$, $R=7$), GLPK required up to $10s$ to compute
the optimal allocation (we used GLPK v4.45 on an AMD Athlon 64 X2
3800+ Dual Core Processor with 4~GB of RAM running Linux 2.6.32). The
auction, implemented as a script in GNU Octave 3.2.3~\cite{octave},
consistently requires less than a second on the same platform. We
recall that the mechanism works on resource-constrained mobile devices
composing DOs; hence, it is important to reduce as much as possible
the overhead to perform the allocation. In practice we should expect
some slowdown due to request-response messages sent through wireless
links.

\begin{figure}[t]
\centering%
\includegraphics[width=\columnwidth]{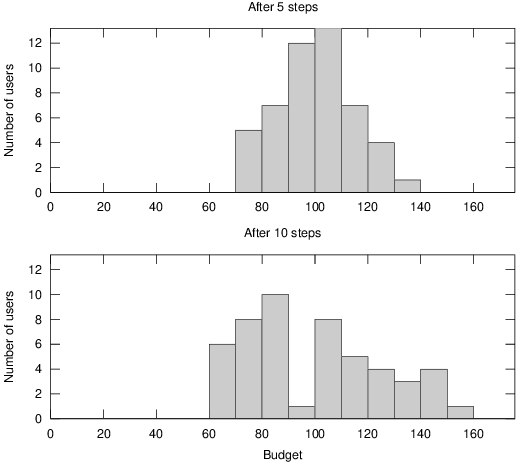}
\caption{Distribution of users budget, after 5 simulation steps (top)
  and at the end of the simulation (bottom); $N=50$,
  $R=7$}\label{fig:hist_budget}
\end{figure}

We show in Figure~\ref{fig:hist_budget} rhe distribution of the users
budget after 5 steps, and at the end of a simulation run (10 steps).
Remember that each user is assigned an initial budget of 100
tokens. The budget distribution spreads over a larger interval as
users trade resources, due to the nature of the experiments carried
out. Each user has the same probability of being a buyer or a seller
at each step as any other user.

\paragraph*{Behavior on crowded markets} 
We also investigated the impact of the connection density (i.e.,
number of links between users) on the total number of matches. We
consider $N=50$ users and $R=7$ resource types, and different values
for the connection density $\rho$ of the ad-hoc network.  The value of
$\rho$ is the fraction of nonzero elements of the adjacency matrix
$M_{ij}$. We considered $\rho=0.2, 0.4, 0.6, 0.8$; for each value, we
executed $T=10$ allocation, and each sequence was independently
repeated $20$ times.

\begin{table}[t]
\centering%
\begin{tabular}{rrrrr}
\toprule
& \multicolumn{2}{c}{Matches} \\
\cmidrule{2-3}
Density & Auction & Optimal & Price & Max Iter. \\
\midrule
0.20 & $103.40 \pm 4.96$ & $113.70 \pm 4.14$ & 1.07 & 6\\
0.40 & $141.00 \pm 3.98$ & $165.60 \pm 3.27$ & 1.15 & 18\\
0.60 & $132.30 \pm 3.21$ & $183.60 \pm 1.82$ & 1.28 & 29\\
0.80 & $89.60 \pm 2.78$ & $190.40 \pm 1.57$ & 1.48 & 56\\
\bottomrule
\end{tabular}
\caption{Number of matches for different connection densities; $N=50$, $R=7$}\label{tab:densities}
\end{table}

\begin{figure}[t]
\centering%
\includegraphics[width=\columnwidth]{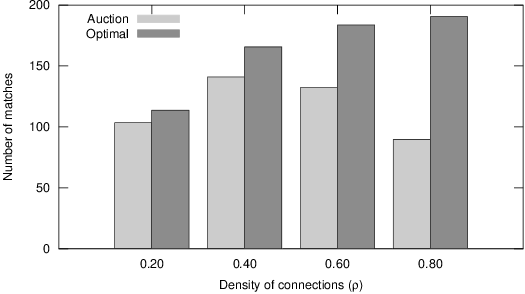}
\caption{Number of matches for different connection densities;
  $N=50$, $R=7$}\label{fig:hist_densities}
\end{figure}

The raw data is shown in Table~\ref{tab:densities}. As the connection
density increases, each buyer can interact with more sellers, since
each node has more neighbors. Therefore, we expect that the total
number of matches increases because a buyer has more chances to find a
seller with enough resources to match his demand.

As we can see from Figure~\ref{fig:hist_densities}, the maximum number
of matches indeed increases as $\rho$ becomes larger. However, the
total number of matches obtained from the auction has a maximum at
about $\rho \in [0.4, 0.6]$, and starts decreasing afterwards. The bad
behavior of the auction-based allocation can be explained by the fact
that for large values of $\rho$, many buyers are likely to share the
same neighboring sellers. Therefore, many buyers are likely also to
share with other buyers the seller offering the lowest price: given
that all buyers will bid the best (lowest) price, this will cause
contention to the ``best'' seller.

\begin{figure}[t]
\centering%
\includegraphics[width=\columnwidth]{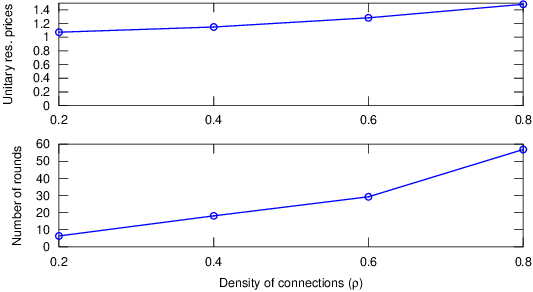}
\caption{Average price for any single resource item (top) and average
  number of rounds to complete an auction (bottom) as a function of
  the connection density}\label{fig:prices}
\end{figure}

To substantiate this claim, we show in Figure~\ref{fig:prices} the
average price for a single resource item, and the average number of
rounds which are necessary for the auction to settle to the final
prices. We observe that both the average resource price and the number
of rounds increases as the connection density $\rho$ becomes larger.
For $\rho=0.8$, about 60 rounds are requested to conclude the
auctions, resulting in higher unitary prices on average.

Several strategies can be used to mitigate the problem below: raising
the prices by a quantity proportional to the excess demand or adding
randomization in the choice of sellers (such that a buyer may
occasionally bid to sub-optimal sellers) are possible extensions which
are currently under investigation. However, we remark that large
connection densities are somewhat unlikely in the scenario we are
considering, as it would require a large number of people sharing
resources within a limited area. And furthermore, in such environments 
short-range communication technologies are preferred for their 
transmission power efficiency.

%%%%%%%%%%%%%%%%%%%%%%%%%%%%%%%%%%%%%%%%%%%%%%%%%%%%%%%%%%%%%%%%%%%%%%%%%%%%%%
\section{Related Works}\label{sec:related}

The development of interaction mechanisms among wireless devices is a
well studied area. For instance, the project ``seamless
computing''~\cite{chalmers2003a} proposed by Microsoft in 2003, and
Jini, which is a part of the Java technology originally developed by
Sun Microsystems, have addressed issues related to the definition and
implementation of ubiquitous computing paradigms.  Today, the notion
of ``digital ecosystem'' is generally used to refer to any distributed
system with properties of adaptability, self-organization, scalability
and sustainability~\cite{Briscoe:2010}. In this paper, we use this
term in a slightly different acceptation, since we give more emphasis
to issues concerned with mobile networks and interaction among
constrained devices~\cite{iwcmc}.

According to our view, there are two kinds of problems which need to
be considered. The first one refers to the distributed resource
utilization. In this sense, several works related to resource
discovery, allocation and organization (mostly in ad-hoc fashion) are
available in the literature,
e.g.~\cite{cooper,Fletcher04unstructuredpeer-to-peer,GruberSK06}.

Another main problem is concerned with optimizing the communication
capabilities of a DO. In general, this issue turns to allow a mobile
node, having multiple wireless network interfaces, to change network
points of attachment (handover) without disrupting existing
connections, combined with the ability to disseminate messages in
multi-hop transmissions (i.e.,~communication in a MANET). Examples of
works on seamless host mobility
are~\cite{Bonola:2009,aict,Mitseva2008,Tsukamoto:2008}

As to the use of auctioning systems to allocate discrete computational
resources, works have been already proposed, but usually employed on
different use case scenarios, such as cluster and classic distributed
systems~\cite{Grosu04auction-basedresource,Ferguson:1996,Kelly04,reddy,stokely}.
Other works employ auction-based mechanisms in wireless networks;
however, usually these are schemes that allow users to dynamically
negotiate their agreed service levels with their service
provider~\cite{Dramitinos05auction-basedresource,TalebN08}, to define
an optimal channel allocation or for scheduling. Hence, it is
something very different from the~\ac{P2P} dynamic resource allocation
we are considering in this work. For example, in~\cite{HuangHCP07},
auction mechanisms are proposed to distributively coordinate and
determine which nodes in a wireless network must act as relay nodes.

%%%%%%%%%%%%%%%%%%%%%%%%%%%%%%%%%%%%%%%%%%%%%%%%%%%%%%%%%%%%%%%%%%%%%%%%%%%%%%
\section{Conclusions}\label{sec:conc}

In this paper we presented a fully distributed algorithm for resource
allocation between DOs. Our algorithm is based on ascending clock
auctions, and allows users to trade resources in exchange for some
form of digital currency. Numerical results show that this approach
represents a viable and effective strategy promoting sharing of
resources, thus providing DOs with the means to form an altruistic
digital ecosystem.

As concerns the general deployment of the proposed scheme in a real
distributed system, there are some open problems that require further
investigation. Security issues are particularly important: for
instance, authentication must be enforced in order to verify the
identity of users that try to utilize resources of other DOs. We will
also consider more general notations to describe requests for
resources, e.g., resource bundles as intervals rather than single
values.

\appendix

\section{Auction Algorithm}\label{app:auction}

Algorithms~\ref{alg:buyer} and~\ref{alg:seller} show the behavior of
the generic buyer $i$ and seller $j$, respectively. We remark that the
same user may act as buyer and seller at the same time, so the
algorithms above could also be executed concurrently by the same node
(for clarity, we neglect synchronization issues in the pseudocode).

\begin{algorithm}[t]
\caption{\textsc{Buyer}$(i)$}\label{alg:buyer}
\begin{algorithmic}[1]
\REQUIRE{$\mathit{RP}_i$ Reserve price of buyer $i$}
\REQUIRE{$\mathit{Req}_{\bullet i}$ Resource bundle requested by $i$}
\STATE{Receive $\mathit{SP}_{\bullet j}, \mathit{Off}_{\bullet j}$ from all $j \in \mathrm{neighbors}(i)$}\label{buyer:rec_sp}
\LOOP
\STATE{$p := 0$}\hfill\COMMENT{Total cost of the bundle}
\STATE{$b_r := 0$, for all $r \in \mathbf{R}$}
\FORALL{$r \in \mathbf{R}$}
\STATE{$S_r := \{ j \in \mathrm{neighbors}(i)\ |\ \mathit{Off}_{rj} \geq \mathit{Req}_{ri} \}$}\label{buyer:S}
\IF{$(S_r = \emptyset)$}
\STATE{\textbf{Abort}}\hfill\COMMENT{Not enough items of $r$ available}
\ENDIF
\STATE{$k := \arg \min_j \{ \mathit{SP}_{rj}\ |\ j \in S_r\}$}\label{buyer:k}
\STATE{$b_r := k$}\hfill\COMMENT{This means $X_{irk} := 1$}
\STATE{$p := p + \mathit{SP}_{r b_r} \times \mathit{Req}_{r i}$}
\IF{$(p > \mathit{RP}_i )$}
\STATE{\textbf{Abort}}\hfill\COMMENT{Reserve price $\mathit{RP}_i$ exceeded}\label{buyer:rp_exceeded}
\ENDIF
\ENDFOR
\STATE{Send bid $\mathit{Req}_{ri}$ to $b_r$, for all $r \in \mathbf{R}$}
\STATE{Receive $\mathit{SP}'_{\bullet j}$ from all $j \in \mathrm{neighbors}(i)$}
\IF{$(\mathit{SP}_{r b_r} = \mathit{SP}'_{r b_r},\ \text{for all}\ r \in \mathbf{R})$}\label{buyer:success}
\STATE{Break}\hfill\COMMENT{Bid successful}
\ENDIF
\STATE{$SP_{\bullet j} := \mathit{SP}'_{\bullet j}$, for all $j \in \mathrm{neighbors}(i)$}
\ENDLOOP
\STATE{Send $\mathit{Req}_{r b_r} \times \mathit{SP}_{r b_r}$ tokens to, and use resource $r$ from seller $b_r$, $r \in \mathbf{R}$}
\end{algorithmic}
\end{algorithm}

Algorithm \textsc{Buyer}$(i)$ requires some additional parameters,
namely the reserve price $\mathit{RP}_i$ of user $i$, and the number
of items of the requested bundle $\mathit{Req}_{\bullet i}$. After
receiving the initial price and offered resource amounts from all
neighbors (line~\ref{buyer:rec_sp}), the main loop starts. At each
iteration, we use variable $p$ to keep track of the cost of the bid;
if $p$ becomes larger than the reserve price $\mathit{RP}_i$, then $i$
resigns and the procedure stops (line~\ref{buyer:rp_exceeded}). We use
the variable $b_r$ to denote the index $j$ of the seller from which
$i$ gets resource $r$.  To place a bid, user $i$ first identifies the
set $S_r$ of neighbors which are offering enough items of resource $r$
(line~\ref{buyer:S}). If $S_r$ is empty, buyer $i$ gives up since we
require that all items in the requested bundle be available. If $S_r$
is not empty, buyer $i$ will bid for resource $r$ to the seller $k \in
S_r$ which advertises the minimum unitary price $\mathit{SP}_{r k}$
for $r$ (line~\ref{buyer:k}).  After all bids are placed, buyer $i$
listens for the new selling prices $\mathit{SP}'_{\bullet j}$ from
each neighbor $j$. If the new prices advertised by sellers to which
$i$ placed a bid match the previous prices,
(line~\ref{buyer:success}), the auction is successful; otherwise, a
new round is performed using the updated selling prices.

\begin{algorithm}[t]
\caption{\textsc{Seller}$(j)$}\label{alg:seller}
\begin{algorithmic}[1]
\REQUIRE{$\mathit{Off}_{\bullet j}$ Resource bundle offered by $j$}
\REQUIRE{$\mathit{SP}_{\bullet j}$ Reserve prices of seller $j$}
\REQUIRE{$\Delta P$ price increment}
\STATE{Send $\mathit{SP}_{\bullet j}, \mathit{Off}_{\bullet j}$ to all $i \in \mathrm{neighbors}(j)$}\label{seller:advertise_SP}
\REPEAT
\STATE{$d_r := 0$ for all $r \in \mathbf{R}$}\hfill\COMMENT{Demand for resource $r$}
\FORALL{bids $\mathit{Req}_{ri}$ received from $i$}
\STATE{$d_r := d_r + \mathit{Req}_{ri}$}
\ENDFOR
\STATE{$\mathit{exc} := \mathit{false}$}
\FORALL{$r \in \mathbf{R}$}
\IF{$( d_r > \mathit{Off}_{rj} )$}\hfill\COMMENT{Excess demand?}
\STATE{$\mathit{SP}_{rj} := \mathit{SP}_{rj} + \Delta P$}\label{seller:incr_SP}
\STATE{$\mathit{exc} := \mathit{true}$}
\ENDIF
\ENDFOR
\STATE{Send $\mathit{SP}_{\bullet j}, \mathit{Off}_{\bullet j}$ to all $i \in \mathrm{neighbors}(j)$}
\UNTIL{$(\mathit{exc} = \mathit{true})$}
\STATE{Receive $\mathit{Req}_{r i} \times \mathit{SP}_{r j}$ tokens and allocate $\mathit{Req}_{ri}$ items of resource $r$ to buyer $i$, $r \in \mathbf{R}$}
\end{algorithmic}
\end{algorithm} 

Algorithm \textsc{Seller}$(j)$ describes the behavior of the generic
seller $j$. The required input parameters are the initial reserve
prices $\mathit{SP}_{r j}$ (the minimum unitary price at which $j$ is
willing to sell resource $r$), the offered quantity
$\mathit{Req}_{rj}$ and the price increment $\Delta P$.  First,
selling prices $\mathit{SP}_{rj}$ are advertised to all neighbors
(line~\ref{seller:advertise_SP}). Then, the main loop starts; we use
the vector $d_r$ to keep track of the total demand of resource $r$, so
that if excess demand for $r$ is detected, its price
$\mathit{SP}_{rj}$ is incremented (line~\ref{seller:incr_SP}). The
loop breaks only when there is no longer excess demand.

\section{Optimization Problem}\label{app:mip}

In this section we formulate the problem of computing the maximum
number of requests that can be satisfied as a~\ac{MIP} optimization
problem. The optimization problem is the following:

\begin{align}
\intertext{Given:}
\mathbf{N} &:= \lefteqn{\{1, \ldots, N\}\ \text{Set of users}}\notag\\
\mathbf{R} &:= \lefteqn{\{1, \ldots, R\}\ \text{Set of resource types}}\notag\\
\mathit{Req}_{ri} &:= \lefteqn{\text{Amount of resource $r$ requested by $i$}}\notag\\
\mathit{Off}_{rj} &:= \lefteqn{\text{Amount of resource $r$ offered by $j$}}\notag\\
M_{ij} &:= \lefteqn{1\ \text{iff user $i$ can interact with user $j$}}\notag\\
\intertext{Define:}
X_{irj} &= \lefteqn{1\ \text{iff user $i$ gets resource $r$ from user $j$}}\notag\\
\intertext{Maximize:} 
&\lefteqn{\sum_{i \in \mathbf{N}} \sum_{r \in \mathbf{R}} \sum_{j \in \mathbf{N}} X_{irj}}\label{eq:maximize}\\
\intertext{Subject to:}
\sum_{i \in \mathbf{N}} \mathit{Req}_{ri} X_{irj} &\leq \mathit{Off}_{rj} && r \in \mathbf{R}, j \in \mathbf{N}\label{eq:const:totaloff}\\
\sum_{j \in \mathbf{N}} X_{irj} &\leq 1 && i \in \mathbf{N}, r \in \mathbf{R}\label{eq:const:unique}\\
\sum_{j \in \mathbf{N}} X_{irj} &= \sum_{j \in \mathbf{N}} X_{i1j} && i \in \mathbf{N}, r \in \mathbf{R}\label{eq:const:all_or_none}\\
X_{irj} &\leq M_{ij} && i \in \mathbf{N}, r \in \mathbf{R}, j \in \mathbf{N}\label{eq:const:neighbors}
\end{align}

The optimization problem above is a~\ac{MIP} problem since it involves
binary decision variables $X_{irj}$.

Constraint~\eqref{eq:const:totaloff} ensures that the total capacity
of each seller $j$ is not violated: the \emph{total} amount of type
$r$ resource provided by $j$ to all other users must not exceed its
capacity $\mathit{Off}_{rj}$. Constraint~\eqref{eq:const:unique}
requires that each user $i$ acquires resource type $r$ from at most a
single provider $j$; note that $i$ may be unable to acquire any
resource at all, so the constraint is an inequality rather than an
equality.  Constraint~\eqref{eq:const:all_or_none} requires that, for
each user $i$, either all the resources it needs are obtained, or none
at all. Finally, constraint~\eqref{eq:const:neighbors} requires that
user $i$ can request resources from user $j$ ($X_{irj} = 1$) only if
$i$ and $j$ are neighbors ($M_{ij} = 1$).

From constraint~\eqref{eq:const:all_or_none} we have that
\[
\sum_{r \in \mathbf{R}} \sum_{j \in \mathbf{N}} X_{irj} = 0\ \text{or}\ R,\quad\text{for all}\ i \in \mathbf{N}
\]
\noindent therefore, the total number of matches, i.e., the total
number of buyers which can be satisfied, is
\begin{equation}
\sum_{i \in \mathbf{N}} \sum_{r \in \mathbf{R}} \sum_{j \in \mathbf{N}} \frac{X_{irj}}{R}\label{eq:maximize_full}
\end{equation}
Since $R$ is a constant, each assignment of $X_{irj}$ maximizing
\eqref{eq:maximize_full} also maximizes the somewhat simpler
expression~\eqref{eq:maximize} which we use as objective function.

\end{document}